\documentclass[fleqn,usenatbib]{mnras}
\usepackage{newtxtext,newtxmath}
\usepackage[T1]{fontenc}
\usepackage{ae,aecompl}
\usepackage{graphicx}	
\usepackage{amsmath}	
\usepackage{amssymb}	
\usepackage{amsmath}	
\usepackage{amssymb}	
\usepackage{csvsimple}
\usepackage{tabularx}
\usepackage{booktabs}

\title[UV upturn at $z=0.7$]{The rise and fall of the UV upturn: $z=0.3,\ 0.55$ and $0.7$}

\author[S.~S. Ali et al.]{S.~S. Ali$^1$\thanks{Email: s.ali@bristol.ac.uk}, M.~N. Bremer$^1$, S. Phillipps$^1$, R. De Propris$^2$ \\
$^1$H.H.~Wills Physics Laboratory, University of Bristol, Tyndall Avenue, Bristol, BS8 1TL, UK. \\
$^2$FINCA, University of Turku, Vesilinnantie 5, 20014 Finland. \\}

\date{Accepted XXX. Received YYY; in original form ZZZ}

\pubyear{2018}

\begin{document}
\label{firstpage}
\pagerange{\pageref{firstpage}--\pageref{lastpage}}
\maketitle

\begin{abstract}
We have analysed the strength of the UV upturn in red sequence galaxies  with luminosities reaching to below the  $L^*$ point within four clusters at $z$ = 0.3, 0.55 \& 0.7. We find that the incidence and strength of the upturn remains constant up to $z=0.55$.  In comparison, the prevalence and strength of the UV upturn is significantly diminished in the $z=0.7$ cluster, implying that the stellar population responsible for the upturn in a typical red sequence galaxy is only just developing at this redshift and is essentially fully-developed by $\sim 1$ Gyr later. Of all the mainstream models that seek to explain the UV upturn phenomenon, it is those that generate the upturn through the presence of a Helium-enhanced  stellar subpopulation on the (hot) horizontal branch that are most consistent with this behaviour. The epoch ($z=0.7$) where the stars responsible for the upturn first evolve from the red giant branch places constraints on their age and chemical abundances. By comparing our results with the prediction made by the YEPS Helium-enhanced spectrophotometic models, we find that a solar metallicity sub-population that displays a consistent upturn between $0<z<0.55$ but then fades by $z=0.7$ would require a Helium abundance of $Y\geqslant0.45$, if formed at $z_f\sim4$. Later formation redshifts and/or higher metallicity would further increase the Helium enhancement required to produce the observed upturn in these clusters and vice versa.
\end{abstract}




\begin{table*}
\begin{filecontents*}{table.csv}
Cluster,z,Filter,Proposal ID,PI,Total exp. time/s
Abell 2744,0.308,ACS F814W,13495,Lotz,104270
Abell 2744,0.308,ACS F606W,13495,Lotz,23632
Abell 2744,0.308,WFC3 F275W,13389,Siana,22720
MACSJ1149+2223,0.543,ACS F814W,13504,Lotz,104177
MACSJ1149+2223,0.543,ACS F606W,13504,Lotz,24816
MACSJ1149+2223,0.543,WFC3 F336W,13389,Siana,22080
MACSJ1149+2223,0.543,WFC3 F275W,13389,Siana,22080
MACSJ0717+3745,0.535,ACS F814W,13498,Lotz,114591
MACSJ0717+3745,0.535,ACS F606W,13498,Lotz,27015
MACSJ0717+3745,0.535,WFC3 F336W,13389,Siana,22720
MACSJ0717+3745,0.535,WFC3 F275W,13389,Siana,22720
SDSSJ1004+4112,0.68,ACS F814W,10509,Kochanek,5360
SDSSJ1004+4112,0.68,ACS F555W,10509,Kochanek,7978
SDSSJ1004+4112,0.68,WFC3 F275W,11732/12324,Kochanek,25914
\end{filecontents*}
\csvautobooktabular{table.csv}
\caption{Table giving details on all the images retrieved from HLA for the four clusters being analysed.}
\label{table_1}
\end{table*}

\section{Introduction}

Despite their high metallicities and large ages, early-type galaxies often show a characteristic
rise in their vacuum ultraviolet flux ($\lambda < 2500$ \AA), a phenomenon dubbed as the UV
upturn or excess (e.g., see review by \citealt{oconnell1999}; upturn henceforth). Although numerous sources were
originally proposed as the origin of the upturn (such as young stars, AGN etc.), energy and morphological considerations favour a population of hot horizontal branch (HB) stars \citep{greggio1990,castellani1991,oconnell1992,
bressan1994}. Such populations have been observed in nearby bulges such as M31 and M32
\citep{brown1998,brown2000} giving rise to their upturns. 

However, the presence of hot HB stars is unexpected in the old and metal rich stellar 
populations that dominate bright early-type galaxies, where the HB should cluster around 
the red giant clump. A metal-poor component would evolve to the blue HB at large ages 
\citep{park1997,yi1998} potentially generating the upturn, but the relative size of the required low metallicity population  would be incompatible with the strength of the metallicity indicators in the optical spectra of bright early type galaxies \citep{bressan1994}. Other possible candidates for the upturn population include 
close contact binaries whose envelope is stripped during the first ascent of the red
giant branch \citep{han2007}; RGB stars with mass loss increasing \citep{yi1997,yi1998,yi1999} as a function of metallicity (needed to 
reproduce the observed correlation between the upturn and Mg$_2$ strength as 
observed by \citealt{burstein1988}) and, finally, a population of metal rich but Helium (He) enriched
stars, that are able to evolve onto the hot HB due to their lower opacity envelopes and
faster evolutionary timescales. This latter hypothesis is particularly attractive
in the light of recent evidence from observations of individual stars in local globular clusters, where anomalously blue
HBs (e.g. as in $\omega$ Centauri and NGC 2808) are now believed to originate from
He-rich subpopulations \citep{norris2004,lee2005,piotto2005,piotto2007}. These
non-cosmological He abundances have now been directly measured by spectroscopy of multiple red giants
in these clusters \citep{dupree2011,pasquini2011,dupree2013,marino2014}, clearly demonstrating the existence of He-enhanced stellar populations. See the two earlier papers in this series \citep{ali2018a,ali2018b}, paper I and II henceforth, for an extended discussion of these mechanisms. 

The evolution of the upturn with redshift should allow us to discriminate between these scenarios. For example, if metal poor stars are responsible, the envelope mass quickly becomes too large for stars to evolve onto the {\it blue} HB, irrespective of their metallicity, only lower mass stars have the potential to evolve onto the blue HB. Consequently the upturn should disappear above some  moderate redshift. Conversely, in the binary scenario, the upturn should always be present, back to the earliest times. However, if metal-rich (and He-rich) stars are responsible, the upturn appears relatively rapidly at moderate redshifts ($z \sim 0.6$ -- \citealt{tantalo1996,chung2017}) and is therefore a sensitive probe of the epoch of galaxy formation and the degree of He-enrichment. Unlike a low metallicity population, the increased He abundance is required if high metallicity stars are to evolve onto the blue HB. Because the progenitors of HB stars have masses such that they take about 8 Gyrs to reach the zero-age line for Helium burning stars, assuming that the responsible population is He-enhanced implies that observations of the upturn at  $z\sim 0.6$ probe conditions in early galaxies at $z > 4$. However, we should note that there is currently no convincing theoretical mechanism that gives rise to the high He abundances in these populations. Nevertheless, there is clear observational proof that such stars exist in the nearby Universe (see review by \citealt{bastian2018} and the Discussion section below).

Most previous work on the redshift dependence of the upturn has been restricted to bright cluster galaxies. In a pioneering series of papers \cite{brown1998b,brown2000b,brown2003} explored the strength of the upturn in a small sample of the most luminous galaxies in three clusters out to $z=0.55$. While the upturn was present out to this redshift, they found some evidence for a difference in the upturn strength above and below $z\sim 0.3$  though with a {\it caveat} that this small sample could suffer from stochasticity. \cite{ree2007} studied a sample of 12 Brightest Cluster Galaxies (BCGs) from Abell clusters at $z<0.2$ and found a fading of the upturn in the past 2~Gyrs. However, a similar analysis by \cite{donahue2010} on 32 BCGs from the Representative XMM-Newton Cluster Structure Survey (REXCESS) between $z=0.06-0.18$ showed no evolution of the upturn over this redshift range. Similarly, \cite{boissier2018} studied a sample of BCGs out to $z=0.35$ behind the Virgo cluster and found that the upturn does not evolve to this redshift. In the local universe, several authors have observed varying levels of upturn strengths in red sequence galaxies (\citealt{boselli2005,smith2012}; paper I), leading to a large scatter in UV--optical colours that is absent in the purely optical colours of the same objects. In paper I we showed that all Coma red sequence galaxies brighter than two magnitudes below $L^*$ have a degree of upturn, that this is not due to star formation, and when modelling the upturn as a blackbody component on top of an old stellar population, its temperature and strength increase with increasing galaxy mass. In paper II we showed that this is also true for galaxies in Abell 1689 at $z=0.18$ where we detect the upturn, with the same characteristics, for the general population of red sequence galaxies to below the $L^*$ point. 

In this paper we extend this analysis to higher redshifts. We measure the upturn in red sequence galaxies for  four more clusters at $z=0.31$ (Abell 2744), $z=0.55$ (MACSJ0717+3745 and MACSJ1149+2223) and $z=0.68$ (SDSSJ1004+4112), again down to the level of typical $\sim L^*$ galaxies. We show that there is no evolution in the prevalence and colour range (i.e. strength) of the upturn population to $z=0.55$ at least in these clusters, but we observe the predicted \citep{bressan1994,tantalo1996} rapid decline at higher redshift; red sequence galaxies in SDSSJ1004+4112 at $z=0.68$ have significantly redder rest-frame $FUV-V$ colours than those of the red sequence population at $z=0.55$. This has important implications for the epoch of galaxy formation and the Helium content of galaxies, which we explore in the discussion. We describe the data in the next section, present our results in section 3 and discuss them in section 4. All magnitudes are quoted in the AB system and we assume the conventional cosmological parameters ($h=0.7, \Omega_M=0.3, \Omega_{tot}=1.0$). Galactic extinction corrections are taken from \cite{schlafly2011} via the NED database.

\begin{figure*}
{\includegraphics[width=0.49\textwidth]{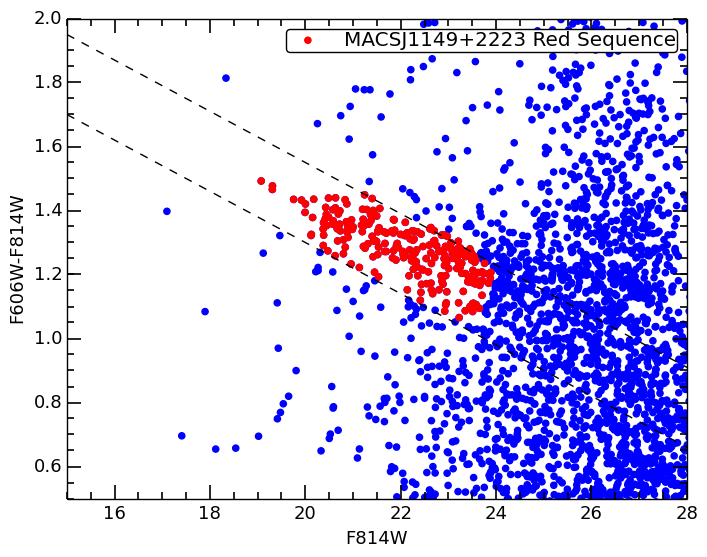}}
{\includegraphics[width=0.49\textwidth]{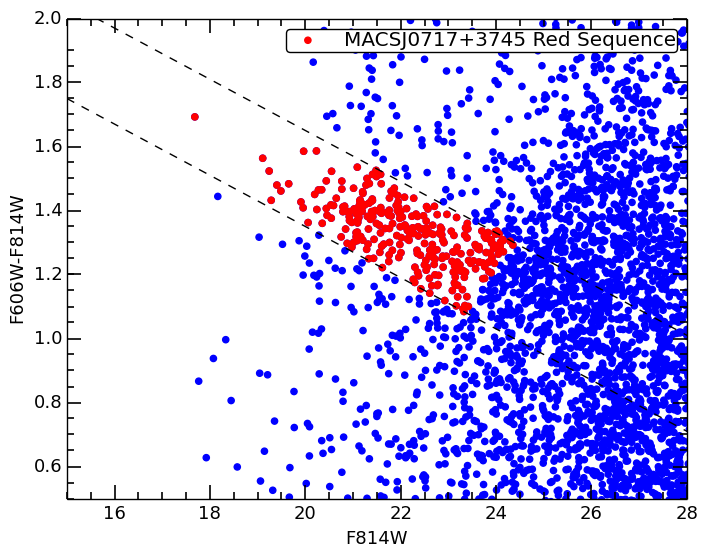}}
{\includegraphics[width=0.49\textwidth]{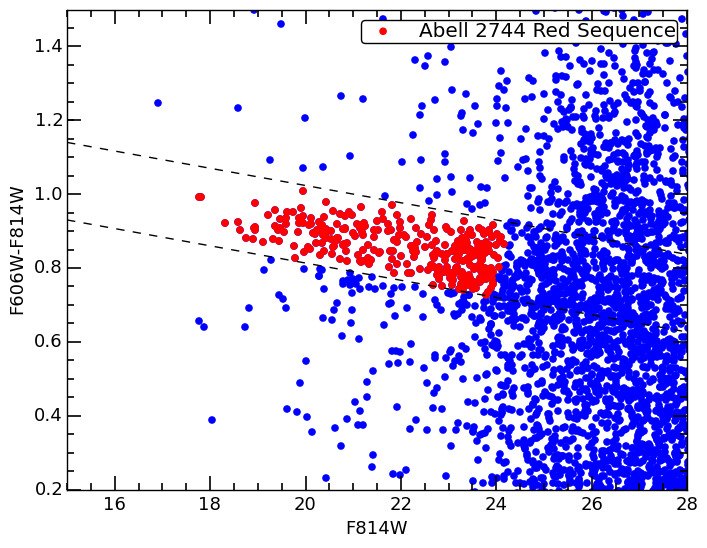}}
{\includegraphics[width=0.49\textwidth]{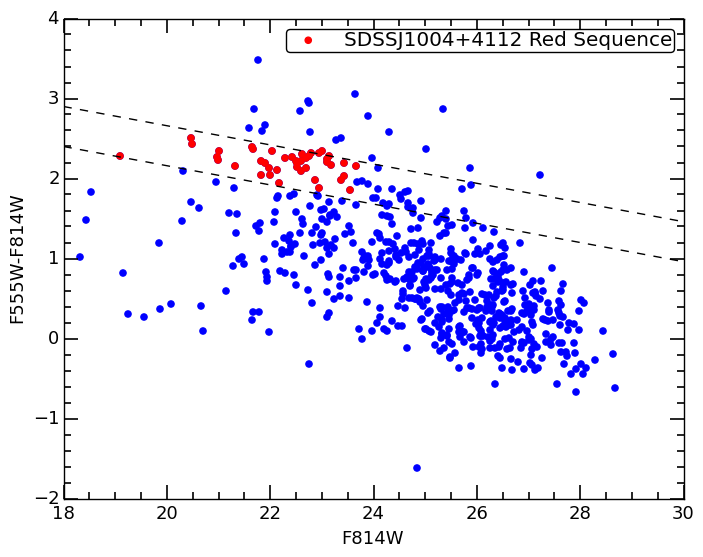}}
\caption{Optical $F606W-F814W$ vs. $F814W$ or $F555W-F814W$ vs. $F814W$ colour-magnitude diagrams for MACSJ1149+2223, MACSJ0717+3745, Abell 2744 and SDSSJ1004+4112. The red sequence is denoted with the red filled circles within the dashed lines and have photometric uncertainties of $<0.05$ magnitudes in their optical colours. The equations of the selection mid-lines are as follows - MACSJ1149+2223: $y=-0.08x+1.825$; MACSJ0717+3745: $y=-0.08x+1.9$; Abell 2744: $y=-0.023x+1.035$; SDSSJ1004+4112: $y=-0.12x+2.65$, where $x$ is the $F814W$ magnitude and $y$ is the relevant colour.}
\label{fig:1}
\end{figure*}

\begin{figure*}
{\includegraphics[width=0.49\textwidth]{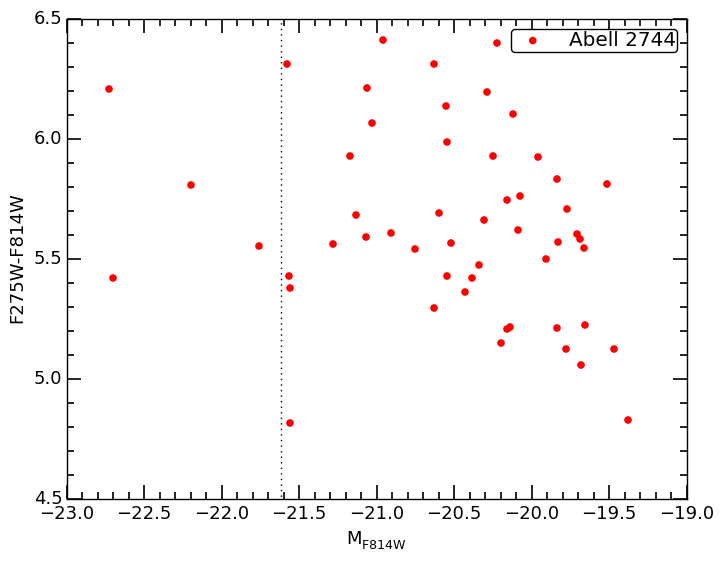}}
{\includegraphics[width=0.49\textwidth]{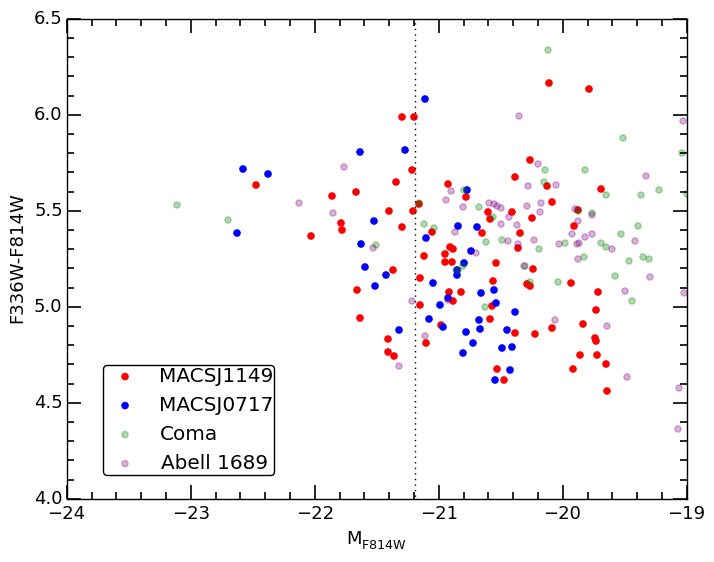}}
\caption{\textit{Left:} $F275W-F814W$ vs. $M_{F814W}$ for Abell 2744. \textit{Right:} $F336W-F814W$ vs. $M_{F814W}$ for MACSJ1149+2223 and MACSJ0717+3745. Also plotted are the same rest frame colours for Coma and Abell 1689 for comparison. The vertical dotted lines in each plot denote the $F814W$ $L^*$ point for each respective cluster \protect\citep{depropris2013}. All photometric errors in colour are $<0.15$ magnitudes unless explicitly plotted. There is no  significant difference in the colour and colour spread of galaxies across the entire redshift range.}
\label{fig:2}
\end{figure*}

\section{Dataset}
For this analysis we required a sample of clusters that had sufficiently deep rest-frame optical and UV data available in the Hubble Legacy Archive.  Only a very limited number of clusters above $z>0.2$ have thus far been observed in suitable UV wavebands by WFC3 to depths that are sufficient to characterise the upturn in red sequence galaxies, and it is this that dictates our choice of clusters. Rest-frame optical images for our clusters were all taken with the Hubble Space Telescope's (HST) Advanced Camera for Surveys (ACS).
Abell 2744 ($z=0.31$), MACSJ0717+3745 and MACSJ1149+2223 (both $z=0.55$) are part of the Hubble Frontier Field sample, providing very deep optical and infrared images for these clusters. We use images in filters F606W and F814W for these clusters, corresponding approximately to rest frame $g$ (4630 \AA) and $r$ (6220 \AA)
for Abell 2744 and to rest frame $B$ (3930 \AA) and $V$ (5270 \AA) for the two $z=0.55$ clusters. For SDSSJ1004+4112 at $z=0.68$, observations were carried out in filters F555W and F814W, corresponding to rest-frame $u$ (3300\AA) and $g$ (4850\AA) respectively. All of these images were retrieved from the Hubble Legacy Archive as fully processed data on which photometry could be performed directly. See Table \ref{table_1} for a full summary of the data, including Proposal IDs, PIs, exposure times, etc.

Ultraviolet data in F275W and F336W were taken for Abell 2744, MACSJ0717+3745 and MACSJ1149+2223 using the Wide Field Camera 3 (WFC3). These correspond to rest frame central wavelengths of $\sim 2100$ \AA\ for Abell 2744 (F275W) and $\sim 1750$ \AA\ and $\sim 2150$ \AA\ for the two $z=0.55$ clusters (F275W \& F336W). For SDSSJ1004+4112 ($z=0.68$) the F275W image corresponds to a rest-frame wavelength of $\sim1650$ \AA. As with the optical data, these images were retrieved as fully processed data from the Hubble Legacy Archive, and multiple frames from each filter were combined together using {\it IRAF}'s {\it imcombine} function, on which photometry was then performed. Refer to Table \ref{table_1} for a full summary of the datasets used. These images have the potential to probe the evolution of the upturn over the past 7 Gyrs. If the upturn arises from a  post main sequence population, this in turn is a probe of star formation at much earlier epochs in these galaxies.

\subsection{Colour-magnitude diagrams}

As in our previous work, we use the red sequence to select quiescent cluster members in all clusters. This has already been shown to be an efficient method to identify early-type galaxies in clusters, with a high degree of fidelity (e.g., see \citealt{rozo2015,depropris2016}). Galaxies belonging to these tight red sequences appear to show no evidence of current or recent star formation, even in high redshift clusters at $z > 1$ (e.g., \citealt{mei2006}). We use SEXTRACTOR \citep{bertin1996} to perform photometry on all optical images listed in Table \ref{table_1} and measure \cite{kron1980} style total magnitudes and aperture magnitudes within a metric diameter of 7.5 kpc. Fig.~\ref{fig:1} shows the optical colour-magnitude diagrams for the clusters, where we identify a tight red sequence of cluster members (see figure caption). All photometrically-selected  objects were then checked by eye to ensure that the final sample only included objects that appeared to have early type morphology, rejecting objects that were clearly (reddened or red) late type galaxies. We select red sequences up to $M_{F814W}\sim24$, as beyond this point the likelihood of contamination from non-cluster foreground or background galaxies increases significantly.

For the selection of the red sequence in the three clusters at $z\leqslant0.55$, a width of $0.2\sim0.3$ in the optical colours was used to be consistent with the red sequence selection of low redshift clusters in our previous papers, where similar rest-frame colours were used in the selection. For Abell~2744, MACSJ1149+2223 and MACSJ0717+3745, we plotted the $F606W-F814W$ vs. $F275W/F336W-F814W$ to check whether there was any correlation between the optical and NUV colours. We found none. This indicates that small changes to the width of the red sequence selection would not significantly affect the spread in the observed UV-optical colours in our clusters. For SDSSJ1004+4112, a slightly larger width of 0.5 magnitudes in $F555W-F814W$ is used since the rest-frame colour corresponds to $u-g$, and it is known that the scatter in the red sequence increases towards shorter wavelengths. Indeed, the scatter in the red sequence in this cluster is the same as that in Coma (\citealt{eisenhardt2007}) for this rest-frame colour.

\begin{figure*}
{\includegraphics[width=0.48\textwidth]{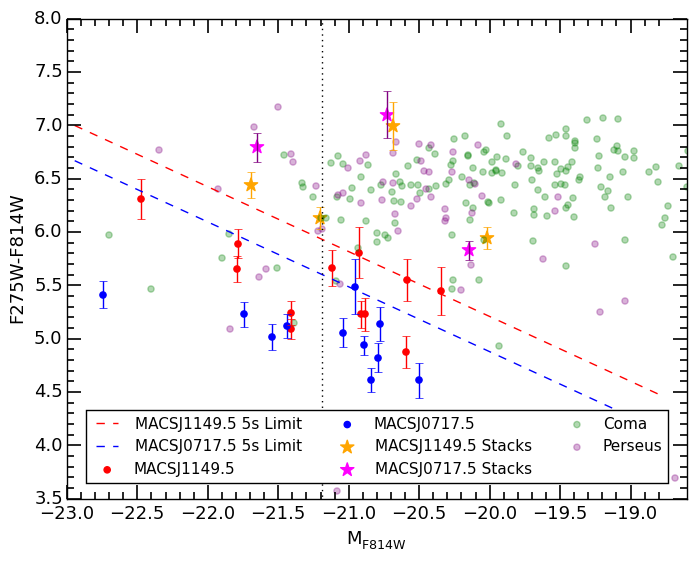}}
{\includegraphics[width=0.48\textwidth]{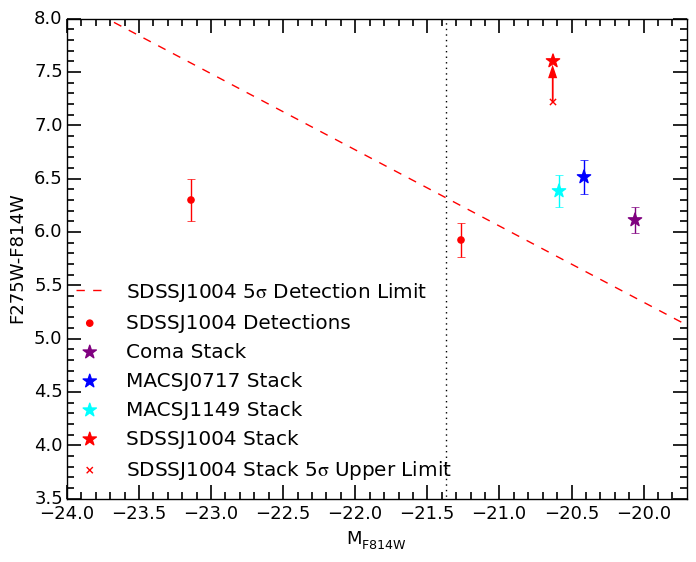}}
\caption{\textit{Left:} $F275W-F814W$ vs. $M_{F814W}$ for MACSJ1149+2223 and MACSJ0717+3745. The magenta and orange starred data points are stacked non-detections from each cluster. Also plotted are the same rest frame colours for Coma and Perseus for comparison. \textit{Right:} $F275W-F814W$ vs $M_{F814W}$ for SDSSJ1004+4112. Also plotted for comparison are the blue, cyan and purple starred points, which represent the stacked colours between $\sim M_{F814W}$=-19.7 to -21.7 for Coma, MACSJ1149+2223 and MACSJ0717+3745. The dashed lines are the UV $5\sigma$ detection limits for each cluster. The vertical dotted lines denote the $F814W$ $L^*$ point for each respective cluster \protect\citep{depropris2013}. The mean colour of galaxies in SDSSJ1004+4112 is at least 1 magnitude redder than at $z\leqslant0.55$, showing that the upturn is either very weak or no longer present at this epoch.}
\label{fig:3}
\end{figure*}

We then measure UV magnitudes for all our red sequence galaxies in each cluster. Given that the UV data were obtained by a different instrument with a different plate scale etc, it was not possible to use SEXTRACTOR in dual image model with the optical data as the master image. We therefore placed metric (7.5 kpc) apertures on the RA and DEC positions of optically-selected red sequence galaxies, after checking the relative alignment of optical and UV images using bright stars. We adopted a 5$\sigma$ cut for detection and also checked all objects by eye to verify that they were really detected in the UV images. In Abell 2744 we detected 53 galaxies in F275W of the 155 red sequence galaxies selected in the optical that were in the field of view of the UV frame. For MACSJ1149+2223 we detected 75 galaxies in F336W and 12 in F275W out of 160, whereas in MACSJ0717+3745 (with higher extinction) these numbers are 39 and 11 in each UV filter, respectively, out of 188 red sequence galaxies. In SDSSJ1004+4112 we detect only 2 galaxies in F275W out of 40 red sequence objects. For all clusters we are able to detect UV emission from galaxies with optical luminosities at least down to $L^*$.

\section{Evolution of the UV upturn}

We first consider the evolution of the upturn using the rest-frame $NUV-V$ colour that we have also used in our previous work (paper I and and II). Although this is not as sensitive to the upturn strength as the standard $FUV-V$ colour, previous studies have shown that $NUV-V$ in red sequence galaxies is mainly produced by the hot HB stars that are recognised as the source of the upturn \citep{schombert2016}. In paper II we have confirmed that the $NUV-V$ colour is indeed mainly produced by hot HB stars and is not influenced heavily by the main sequence age and metallicity of the underlying stellar population in quiescent galaxies. This is further developed below in the discussion. For our clusters, the rest-frame $NUV$ lies within the F275W (Abell 2744) and F336W (MACSJ0717+3745 and MACSJ1149+2223) filters, which have greater sensitivity than bluer HST filters closer to the original $FUV$ filter. For SDSSJ1004+4112, the F275W image corresponds more  closely to FUV than the NUV in the rest-frame.

\subsection{NUV-optical}

Fig.~\ref{fig:2} (left) shows the F275W--F814W (approximately $NUV-r$) against $M_{F814W}$ colour-magnitude diagram for red sequence galaxies in Abell 2744 ($z=0.31$). We find a spread of 1.5 magnitudes in $NUV-r$ colour, which is perfectly consistent with that measured in Coma and Abell 1689 in paper I and II. This suggests that there has been no evolution of the upturn strength at least to this redshift.

We plot reddening-corrected observed F336W--F814W colours vs. $M_{F814W}$ for the two $z=0.55$ clusters in Fig.~\ref{fig:2} (right), which is equivalent to $NUV-V$. For comparison, we also plot the same colours in Coma and Abell 1689 in the same absolute magnitude range as sampled in MACSJ0717+3745 and MACSJ1149+2223. We find that these objects have the same range in colour $4.5 < NUV-V < 6$ at all redshifts. This indicates that at least to $z=0.55$ the upturn has not evolved significantly. By itself, this result rules out a low metallicity hot HB population as the origin for the upturn, as such stars would only reach the hot HB at significantly lower redshift -- they should not be present in $z=0.55$ populations.

\subsection{FUV-optical}

For MACSJ0717+3745 and MACSJ1149+2223 we have archival HST imaging in the F275W filter, corresponding to rest-frame 1750 \AA\ (at $z=0.55$), close to the classical definition of the FUV band used to characterise the upturn. We plot the observed, reddening-corrected F275W--F814W vs. $M_{F814W}$ for both clusters in Fig.~\ref{fig:3} (left), which is equivalent to rest-frame $m_{1750}-V$. 

In MACSJ0717+3745 and MACSJ1149+2223 we are only able to clearly detect (at least $5\sigma$ and visually confirmed)  the galaxies with the strongest upturns. Recall that at any $V$ magnitude the spread in the upturn colour is of the order of 1.5$\sim$2 mags, resulting in 12 and 11 objects in each cluster, respectively. This means we are not detecting the entire population of galaxies exhibiting upturns because of our detection limit -- i.e. a galaxy with no upturn would have a UV magnitude significantly below our detection limit, as shown in Fig.~\ref{fig:3} (left). In this figure we also show the equivalent colours for Coma and Perseus, shifted to $z=0.55$ and bandpass-corrected to the F275W filter, following \cite{chung2017}. Only the strongest upturn galaxies from Coma and Perseus would lie above the detection limit at this redshift.

\begin{figure}
\includegraphics[width=0.50\textwidth]{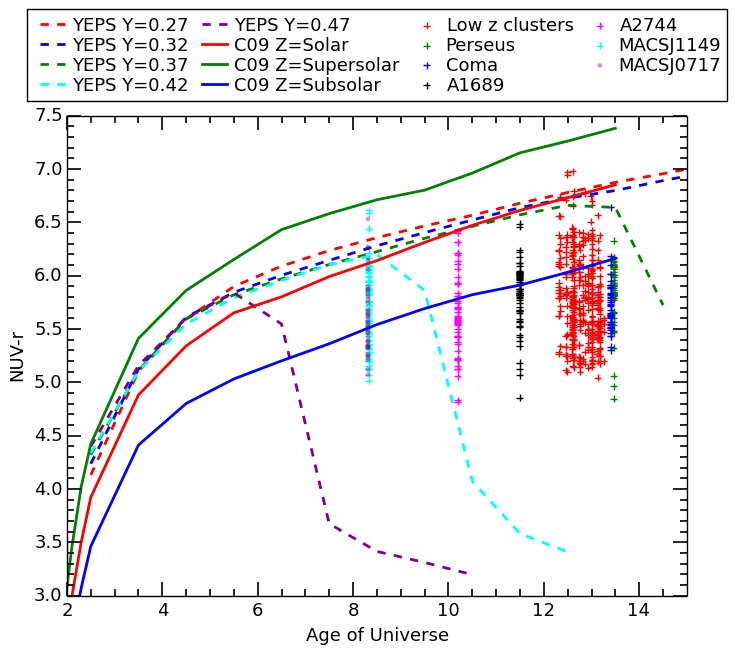}
\caption{YEPS spectrophotometric models (assuming $z_f=4$ and $Z=Z_{\odot}$) showing the evolution of the GALEX $NUV-r$ colour over the age of the universe for a range of Helium abundances. Also included in the plot are the predictions from the C09 model with the following metallicities: $Z$=\(Z_\odot\), 0.56\(Z_\odot\) and 1.78\(Z_\odot\). Plotted on top are the $NUV-r$ colours of Coma, Perseus, Abell 1689, Abell 2744, MACSJ1149+2223, MACSJ0717+3745 and 24 other clusters between $0<z<0.1$ as detailed in the text. Photometric uncertainties in colours are <0.15 magnitudes. The upturn is detected at least to $z=0.55$ with unchanged strength for galaxies down to the $\sim L^*$ level.}
\label{fig:yeps}
\end{figure}

In order to explore the rest of the upturn population in these clusters, we produced mean stacks of the non UV-detected red sequence galaxies in these clusters in the F275W filter. For each galaxy we extracted a $8''$ by $8''$ cutout from the HST image, at the RA and DEC position established from the F814W image. Stacking the 5 optically brightest galaxies lacking individual UV detections (to our limit) in MACSJ1149+2223 and the 7 optically brightest UV non-detections in MACSJ0717+3745 achieves a $>5\sigma$ detection in the stacked F275W data in both cases. We then made further stacks from optically fainter sources by combining as many galaxies as necessary (in order of F814W magnitude) to produce stacks with approximately the same combined F814W flux as the first stack. We repeated this until we reached galaxies as faint as  $M_{F814W}\approx-19.5$ (equivalent to the limits previously reached in Coma, Fornax and the compilation of 24 2dFGRS clusters we describe in the discussion). This resulted in  a total of 4 stacks for MACSJ1149+2223 made up of 5, 8, 15 and 26 galaxies and 3 stacks for MACSJ0717+3745 consisting of 7, 18 and 32 galaxies. This procedure means that if galaxies of different optical brightnesses had, on average, the same UV-optical colours, each stack should be detected in the UV at the same level, although given the numbers in each stack, clearly there would be stochastic variation due to small number statistics. We plot the F275W-F814W colour of these stacks against F814W, where the F814W used is the mean  magnitude of the galaxies contributing to that stack,  in Fig.~\ref{fig:3} (left). The range of colour seen in these stacks is similar to that seen in the range of individual colours for upturn systems drawn from Coma and Perseus, $5.5 < FUV-V < 7$. In combination with the small number of individual detections, this behaviour implies that the upturn has undergone no significant evolution to $z=0.55$ (at least for these clusters), as already established from the $NUV-V$ data presented in the previous subsection and shown in Figs. \ref{fig:2} and \ref{fig:yeps}.

\begin{figure}
\includegraphics[width=0.5\textwidth]{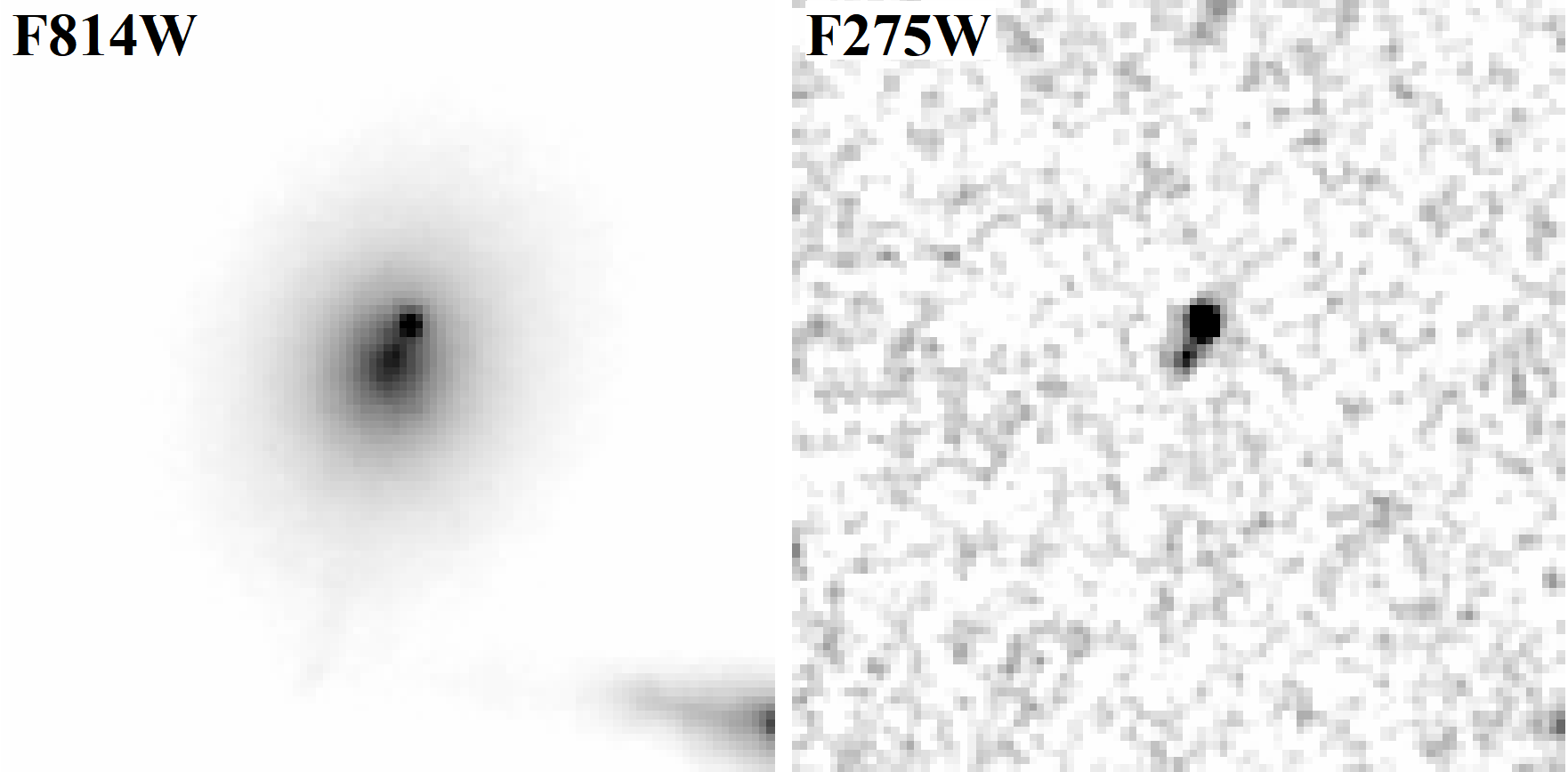}
\caption{5'' by 5'' and 4'' by 4'' images of the SDSSJ1004+4112 BCG in the F814W (optical) and F275W (UV) band respectively. The bright point-like source to the north-east of the BCG in F814W is a lensed quasar at $z=1.74$. The quasar is much bluer than the BCG in $F275W-F814W$ and this affects the quality of the latter's photometry in $F275W$. This is reflected in the uncertainty in the applied error bar to the photometry of the BCG.}
\label{fig:4}
\end{figure}

\subsubsection{SDSSJ1004+4112: The Rise of the Upturn}

For SDSSJ1004+4112 we used archival HST imaging in the F275W filter, corresponding to rest-frame 1650\AA\ (at $z=0.68$). In Fig.~\ref{fig:3} (right), we plot F275W--F814W against F814W, corresponding to rest-frame $m_{1650}-g$. Brighter than $L^*$, only one of three galaxies meeting our optical cut is individually detected in the UV at $>5\sigma$, which is the brightest cluster galaxy. However, photometry for this object is complicated by the presence of a $z=1.47$ lensed quasar image projected close to the centre of the galaxy \citep{inada2003}. Both the quasar and the galaxy are detected in the UV, with the  brightest emission arising from the point source (Fig. \ref{fig:4}). Consequently, for this object we calculated the optical and UV fluxes in apertures that excluded the region contaminated by the quasar, so the brightest data point in Fig. \ref{fig:3} (right) is the colour of the central region of the BCG with minimal contamination from the quasar. We estimate the uncertainty on this value to be approximately 0.2 mags. Besides the BCG, we have one other galaxy (at $\sim L^*$) detected at a $>5\sigma$ level in the UV. The detections are shown as filled red circles in Fig. \ref{fig:3} (right).

\begin{figure}
\includegraphics[width=0.5\textwidth]{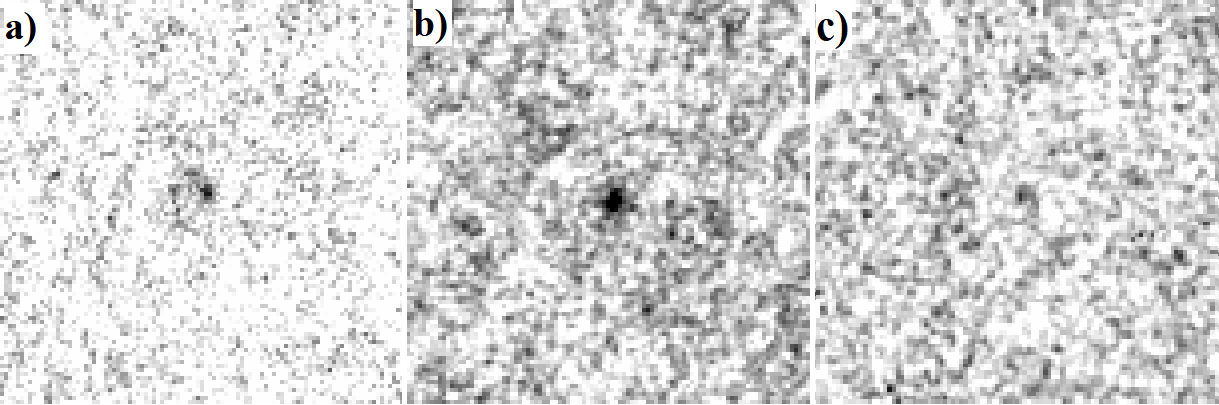}
\caption{4'' by 4'' stacks of all galaxies between $M_{F814W}$=-19.7 to -21.7 in MACSJ1149+2223 (a), MACSJ0717+3745 (b) \& SDSSJ1004+4112 (c) in the F275W band. The stacks of MACSJ1149+2223 and MACSJ0717+3745 both demonstrate $>5\sigma$ detections, while the SDSSJ1004+4112 stack corresponds to a $\sim3\sigma$ detection.}
\label{fig:5}
\end{figure}

At fainter optical magnitudes we have no individual UV detections, but given the 5$\sigma$ limit shown in Fig. \ref{fig:3} (right), we may not expect there to be any if emission arises from upturns of comparable strength to those seen at lower redshift.  Given this, we can only search for a general upturn population by carrying out a stacking procedure on the fainter red sequence galaxies in this cluster, similar to that used in the analysis of  the $z=0.55$ systems. In this case, we simply stacked the optical and UV images from the 20 red sequence galaxies with optical magnitudes in the range $-19.7 < M_{F814W} < -21.7$ (excluding the one $>5\sigma$ detection in this interval), a range similar to that explored in the analysis of the lower redshift clusters. This is therefore representative of the general population of objects contributing to the upturn in local clusters, without including fainter dwarfs whose low metallicity may produce significant flux in the UV from lower metallicity old stellar populations. No detection is observed in the stacked UV image to a 5$\sigma$ {\sc upper limit} to optical-UV colour of $F275W-F814W>7.2$. A possible $\sim 3\sigma$ detection may be present (Fig. \ref{fig:5} with a colour of $F275W-F814W=7.6$ for the stacked flux from the 20 galaxies. Both the $5\sigma$ upper limit and the potential $3\sigma$ detection are plotted in Fig. \ref{fig:3} (right). 
 
As is apparent from Fig. \ref{fig:3} (left), the 5$\sigma$ limit to the stacked colour is as red or redder than the individual  colours of Coma and Perseus red sequence galaxies in the same rest-frame bands to the same luminosity limit. To further understand the significance of this, we combined the fluxes in both  the UV and optical of all galaxies in Coma within the same $g$ band luminosity range as SDSSJ1004+4112 (corrected for passive evolution using the fiducial model by \citealt{conroy2009}) and plot the resulting equivalent F275W-F814W colour in Fig.~\ref{fig:3} (right). The resultant colour is a magnitude bluer than the 5$\sigma$ upper limit and $\sim 1.5$ magnitudes bluer than the possible detection for the $z=0.68$ stack (see Fig. \ref{fig:5}). We also repeat this process for MACSJ0717+3745 and MACSJ1149+2223, stacking only non-detections (to avoid the biased sample of detected objects in these clusters, that must be among the galaxies with the stronger upturns). These stacks have clear detections at the 5$\sigma$ level, the values of  which are also plotted in Fig.~\ref{fig:3} (note that we have made small $k$-corrections to the colours to convert them to the equivalent of $z=0.68$ and F275W--F814W). These stacks are again bluer than the upper limit for the stacked galaxies in SDSSJ1004+4112. The stacked images are displayed in Fig. \ref{fig:5}. Barring the two individually-detected objects in the $z=0.68$ cluster, one of which is the BCG, the upturn appears to have significantly faded in the general cluster red sequence population between $z=0.55$ and $z\sim0.7$ in a relatively rapid fashion, as predicted by the high metallicity (and He-enhanced) models of \cite{bressan1994} and \cite{tantalo1996}.

Additionally, we note that MACSJ0717+3745 suffers from significantly higher Galactic extinction than the other clusters (about 0.4 mag. in F275W). By chance, the effect of this extra dust extinction  nearly compensates for the higher distance modulus of SDSSJ1004+4112, especially given that the two images have similar exposure times of about 25ks. If both systems had similar upturn populations, we might then expect to see similar detection statistics for these populations. Despite this, we obviously detect more galaxies to a similar observed UV detection in MACSJ0717+3745 than in SDSSJ1004+4112,  and in stacking the non detections we generate  clear detections in F275W, at a level that is significantly bluer than the upper limit to the non-detection of the  equivalent stack in SDSSJ1004+4112. This is further evidence that the upturn has significantly faded across this redshift range - if it didn't we would have expected the two raw UV data sets to be more similar than they are. 

We should note that we rejected several galaxies brighter than $M_{F814W}=-21$ on morphological grounds that made the initial optical colour cut for the red sequence in SDSSJ1004+4112. One appeared to be an interacting system and others appeared to have disks that were too structured  or edge-on systems too thin to be S0 galaxies. However, these could eventually evolve into classical lower red sequence galaxies (e.g. \citealt{depropris2016}) and so we might be excluding a section of the cluster galaxy population  that exhibit upturn populations at these redshifts. However, none of these are individually detected at the $5\sigma$ level in the UV and so it is unlikely that these systems exhibit upturns of a strength comparable to lower redshift cluster red sequence galaxies of comparable optical luminosity.

Our analysis of this whole sample of clusters out to $z\sim 0.7$ therefore implies  that the upturn arose  across a $\sim 1$~Gyr period between $z\sim 0.7$ and $z\sim0.5$ in the general population of galaxies down to $\sim L^*$ in the optical. In the next section we explore the consequences of our findings for galaxy formation and evolution.

\begin{figure*}
\includegraphics[width=0.49\textwidth]{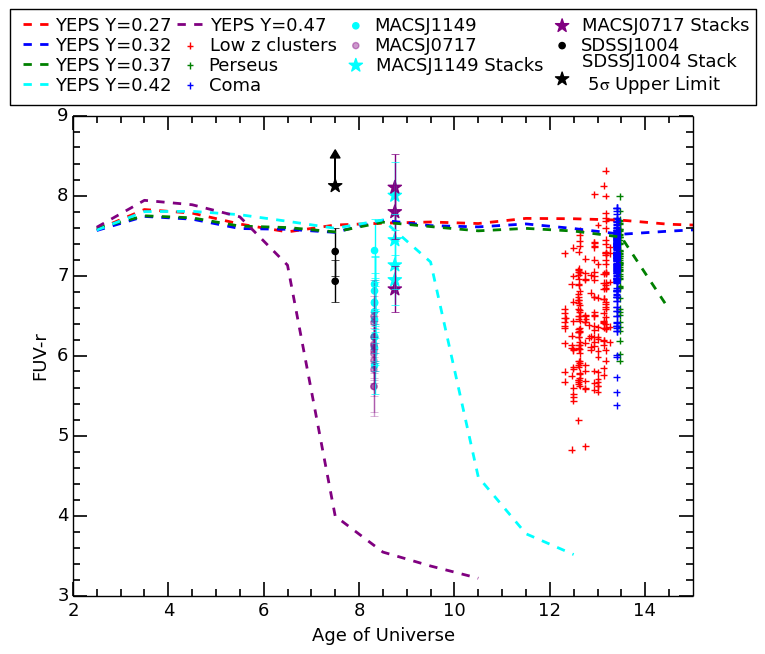}
\includegraphics[width=0.49\textwidth]{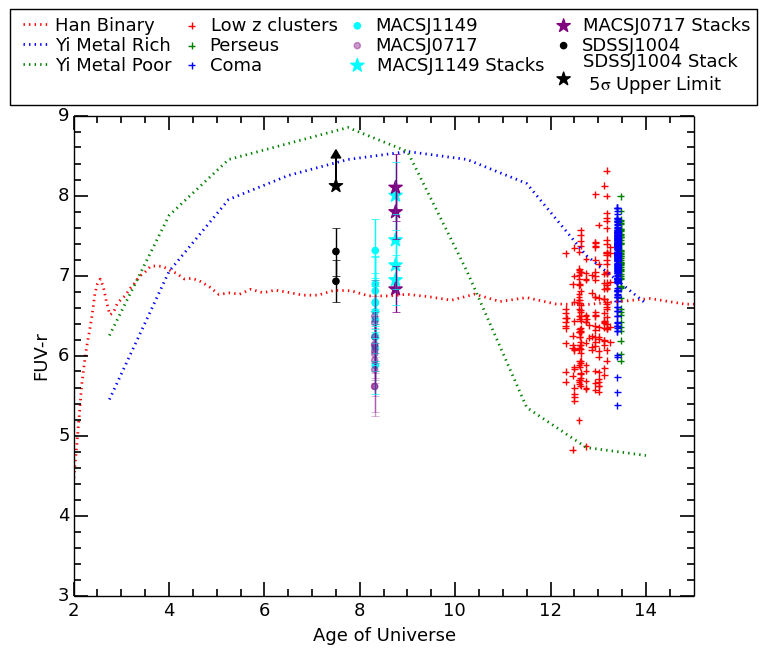}
\caption{{\it{Left}}: YEPS spectrophotometric models (assuming $z_f=4$ and $Z=Z_{\odot}$) showing the evolution of the GALEX $FUV-r$ colour over the age of the Universe for a range of Helium abundances. Plotted on top are the $FUV-r$ colours of Coma, Perseus, a local sample of
clusters, MACSJ1149+2223, MACSJ0717+3745 and SDSSJ1004+4112 with appropriate k-corrections as detailed in the text. For the $z=0.55$ clusters, both the individual detections and the stacks (shifted slightly to the right for visual clarity) are plotted from Fig. \ref{fig:3}. {\it{Right}}: Same photometric results but this time plotted against the $FUV-r$ colours from the binary model of \protect\cite{han2007}, as well as the metal-rich and metal-poor models of \protect\cite{yi1999} (Model A and C from their Fig. 7).} Photometric uncertainties in colours are $<0.15$ magnitudes unless explicitly plotted. The detection of a decrease in the strength of the upturn at $z > 0.65$ implies that a metal-rich He-rich subpopulation with $Y > 0.42$ and $z_{f}\geqslant4$ is present in cluster early-type galaxies at a mass fraction of $\sim 10\%$ and is inconsistent with other proposed sources (metal poor stars, close binaries or mass loss).
\label{fig:yeps_fuv}
\end{figure*}

\section{Discussion}

\subsection{Helium rich stars as the source of the UV upturn}

Our results demonstrate that the range and strength of the upturns in typical cluster red sequence galaxies remain constant between $z=0$ and $z=0.55$. The $FUV-V$ and/or $NUV-V$ colours (a measure of the upturn strength) between low redshift clusters such as Coma and Perseus are nearly identical to those of Abell 2744 ($z=0.308$) as well as MACSJ1149+2223 and MACSJ0717+3745 ($z=0.55$). Assuming a hot HB origin for the upturn, this phenomenon should show an eventual decline with increasing redshift that corresponds to the age at which a blue HB can first start to form, and as such the upturn has not had enough time since the galaxy formed to develop from stars evolving off of the RGB. The upturn should develop rapidly once the turnoff mass reaches a point that allows for the HB to form (\citealt{tantalo1996}). In our analysis of SDSSJ1004+4112, we find that the number of galaxies demonstrating an upturn of similar strength as cluster galaxies at $z\leqslant0.55$ decreases significantly by $z=0.68$ - all but two of the brightest galaxies show little to no detectable upturn in rest-frame $1650-g$ colour at this redshift even after stacking galaxies in a bin spanning 2 magnitudes around the $L^*$ point. Stacking galaxies in the same magnitude range in Coma, MACSJ1149+2223 and MACJ0717+3745 demonstrate strong upturns in each case as discussed earlier. Particularly strong evidence for the fading comes from the fact that MACSJ0717+3745 has an extinction in its $F275W$ band of approximately 0.4 magnitudes, which compensates for the difference in the distance modulus between $z=0.55$ and $z=0.68$, thus making the depth of the data effectively identical to that of SDSSJ1004+4112 in the UV. Despite this, clear detections are made in the UV with and without stacking in MACSJ0717+3745, while very few galaxies with a clear upturn detection are seen in SDSSJ1004+4112, despite SDSSJ1004+4112 having a longer total exposure time in $F275W$ compared to MACSJ0717+3745. These results indicate that for typical $\sim L^*$ cluster red sequence galaxies, the upturn emerges at around $z=0.7$. The individual detection of two of the most massive galaxies (including the BCG) may indicate that their upturns develop earlier, possibly because these are the galaxies with the oldest stellar populations in the cluster. This is consistent with the results of \cite{lecras2016} who identified evidence for upturns out to $z\sim 1$ in some (but not all) of the most massive (stellar masses above $10^{11.5}$M$_\odot$) BOSS early-type galaxies. Alternatively, we could simply be seeing the impact of stochasticity as the upturn starts to develop over a limited range in time across the cluster's galaxy population. Regardless, the upturn must then be fully developed in typical red sequence cluster galaxies  by $z=0.55$ in order to explain our observations in MACSJ1149+2223 and MACSJ0717+3745. The difference in lookback time between these two redshifts is $\sim1$ Gyr, a timescale which matches the predictions made by \cite{tantalo1996} and \cite{chung2017} for the rapid appearance of an upturn sub-population once the hot HB starts to form.

Our data to $z=0.55$ are broadly consistent with previous work by \cite{brown1998b,brown2000b,brown2003} on the brightest  galaxies in clusters in the sense that both sets of studies detect upturns in cluster populations out to this redshift. \cite{lecras2016} use a series of indices (developed by \citealt{fanelli1992}) to measure the strength of the hot HB population in BOSS Luminous Red Galaxies and show evidence for a decline in the upturn strength at $z > 0.6$, though with a subset of the most massive galaxies  showing an upturn even at $z=1$. A caveat on this is  that some of the indices used in their work also have some sensitivity to star formation, and \cite{roseboom2006} has shown that there is a large fraction of post-starburst galaxies in the LRG sample.

Given the rapid onset of the upturn at $z\sim0.7$ which then remains consistent up to present day with a range of upturn strengths, our observations can be best explained through the presence of a He-enhanced sub-population of HB stars (superimposed on top of the majority `red and dead' population that dominates early-type galaxies) for which a transition in the upturn strength is expected at a moderate redshift, as will be shown below. We will also discuss why other proposed sources of the upturn cannot fully explain our observations.

\begin{figure*}
\includegraphics[width=0.49\textwidth]{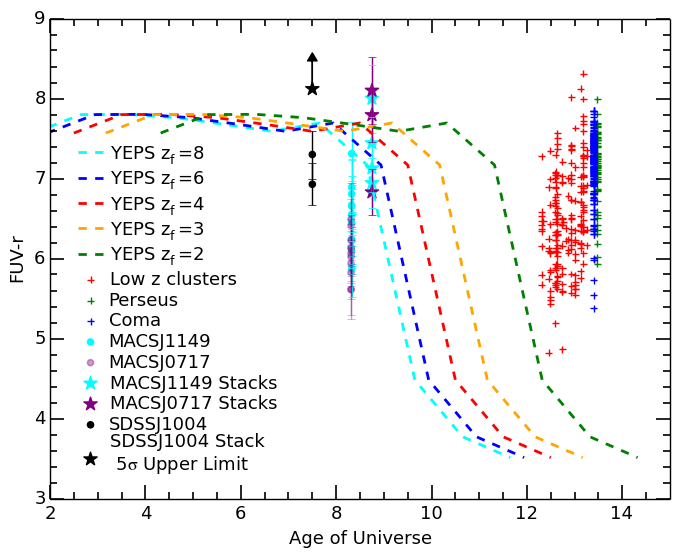}
\includegraphics[width=0.49\textwidth]{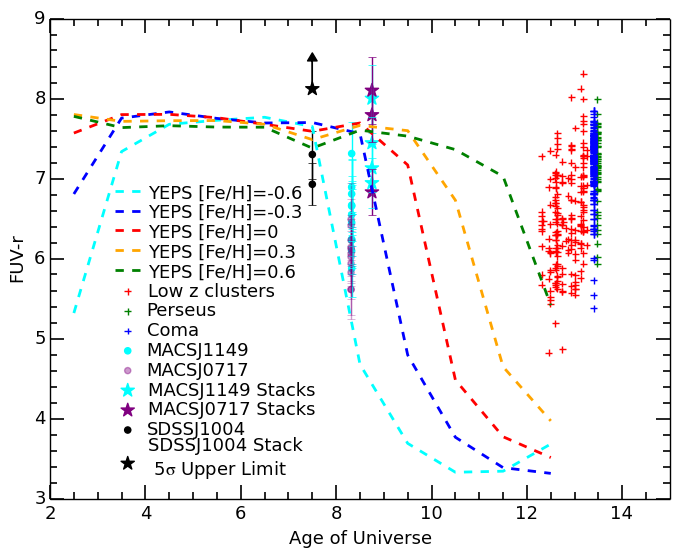}
\caption{YEPS spectrophotometric model for $Y=0.42$ (from Fig. \ref{fig:yeps_fuv} as an example with the same photometric results) showing the evolution of the GALEX $FUV-r$ colour over the age of the Universe for a range of formation redshifts and metallicities. These plots show that for a given $Y$, later formation redshifts or increasing the metallicity delays the onset of the upturn and vice versa.}
\label{fig:8}
\end{figure*}

\subsection{The ages and Helium abundances of galaxies}

Assuming a He-enhanced HB origin for the upturn, it is then possible to interpret our results within the framework of stellar population synthesis models which allow for a variation in He-abundance, specifically the YEPS models \citep{chung2017}. These  are a series of Simple Stellar Populations that evolve with age depending on the He-abundance ($Y$) and metallicity ($Z$). These allow us to make   predictions on the age and $Y$ of the sub-populations which can  give rise to the upturn in our galaxies. In doing so, we  note here that \cite{chung2017} tabulate the YEPS models for a range of $Z$ and $Y_{ini}$, defined as the initial Helium content of an SSP for which notionally $Z=0$. The He-abundance, $Y$, in an actual model is then related to these parameters by $Y=Y_{ini}+2Z$. We use this relation  to calculate $Y$ for our models and use these values in the discussion hereafter.

Fig. \ref{fig:yeps} shows the evolution of the GALEX $NUV-r$ colour against the age of the Universe as predicted by the YEPS models with a range of $Y$ values. We assume $Z$=\(Z_\odot\) and a redshift of formation for the stellar population of $z_f$=4. We also plot the time evolution of the same colour for  models from \cite{conroy2009}, C09 henceforth,  for three different metallicities: $Z$=\(Z_\odot\), 0.56\(Z_\odot\) and 1.78\(Z_\odot\), i.e. solar, sub-solar and super-solar. Once again we assume a $z_f$=4 for all models. The C09 models have no He-enhancement and as such  have no contribution from a hot HB sub-population. Their $NUV-r$ colours are purely driven by the passive evolution of a  canonical old stellar population that is thought to be the dominant stellar population in  early-types galaxies. The solar metallicity C09 model evolves  nearly identically  to the $Y=0.27$ YEPS model, because  the latter does not develop an upturn by the present day, unlike the higher $Y$ models. Consequently, neither of these models show the characteristic increase in the UV flux at late ages  expected from a He-enhanced population. The YEPS models are for a population which has a single value of $Y$, i.e. the predicted colours are for a stellar population where {\it all} stars have the same He-abundance. As shown in paper I it is likely that cluster red sequence galaxies, even those with the strongest upturns, can be modelled with a He-enhanced population of a few percent of the total, the rest being the conventional stellar population of the type modelled in C09. Consequently, we should not expect to see colours as extreme as the bluest YEPS model predictions. The $\sim 3$~mag difference in the bluest observed $NUV-r$ and YEPS predictions is consistent with a He-enhanced population of up to of order $\sim 10$ percent of the total stellar population in these galaxies. The YEPS models are tabulated for five  discrete values of $Y$ at each $Z$ and so in the following we linearly interpolate between these values of $Y$ where necessary.

Also plotted in Fig. \ref{fig:yeps} are the rest-frame $NUV-r$ colours of Coma and Perseus cluster galaxies from paper I, those for the Abell~1689 population from paper II and the red sequence galaxies from 24 2dFGRS clusters at $z<0.12$ (as described in \citealt{depropris2017}). Finally we plot the rest-frame $NUV-r$ colours of the red sequence populations of  Abell 2744, MACSJ1149+2223 and MACSJ0717+3745 from this paper, allowing us to compare the evolution of the observed upturn between $z=0$ and $z=0.55$ in a large range of cluster galaxies with the predictions from the models. For the $z=0.55$ clusters, where the observed optical band corresponds to the rest-frame $V-$band, the optical photometry was transformed to $r$ through a k-correction of $V-r=0.4$ derived from C09, to match the $NUV-r$ colour of the models.

From first glance one can see that in all clusters out to $z=0.55$  the $NUV-r$ colours range   between $\sim5-6.5$ in each case. The clear implication is that the  range and strengths of upturns have not evolved in any significant manner over this redshift range. Comparing our results to the YEPS models, we find that most of the reddest galaxies in our sample match the $NUV-r$ colours of the solar metallicity C09 model and the YEPS $Y=0.27$ model, i.e. the models with little or no upturn. The few galaxies that have redder colours than these models can be easily accommodated with a slightly super-solar metallicity model. This is expected as the majority of galaxies in our sample are giant ellipticals and S0s, which are very likely to have solar or super-solar metallicities (see \citealt{price2011} for an analysis of the metallicity in Coma galaxies). Alternatively, the galaxies in our sample with the bluest $NUV-r$ colours constrain the parameters of valid YEPS models through the onset of the hot HB. The bluest galaxies at $z=0.55$ constrain any He-enhanced model to have $Y>0.42$ and, if we interpolate between the tabulated values plotted in Fig. \ref{fig:yeps}, $Y \geqslant0.45$ provided that $z_f$=4 or lower. Although the constraint on $Y$ can be reduced by increasing $z_f$, because $z=4$ is only about 1~Gyr after the  first  galaxies are thought to have formed, any relaxation still requires a significant He-enhancement. 

This prediction for $Y$ is further reinforced by the observed fading in the rest-frame $1650-g$ colour in SDSSJ1004+4112. Note that data points for SDSSJ1004+4112 could not be directly plotted in Fig. \ref{fig:yeps} as the rest-frame UV data for this system probes the region around 1650\AA\ and any transformation to  rest-frame $NUV$ requires  a large and, crucially,  model dependent k-correction. Galaxies containing a few percent of stars with  $Y\sim 0.45$  would be  relatively red in the $NUV-r$ at $z=0.68$ (age of Universe $\sim7.5$ Gyrs) because the He-enhanced population would have colours barely different from the majority population at this point,  but then get rapidly bluer over the next Gyr, so that by  $z=0.55$ the most extreme members of the population  would be blue enough to cover the entire range of $NUV-r$ colours seen in MACSJ1149+2223 and MACSJ0717+3745. As noted above, an earlier formation redshift can be traded for slightly less He-enrichment ($Y=0.43\sim0.44$ at $z_f$=6-8), but studies by \cite{jorgensen2017} suggest formation redshifts between 2<$z_f$<6 for cluster galaxies.

Fig. \ref{fig:yeps_fuv} (left) is a similar plot, but this time for the $FUV-r$ colours predicted by the YEPS models and the observed photometry in all of our clusters with appropriate data, including the values for the photometry of the stacks displayed in Fig. \ref{fig:3} (left) and the upper limit from Fig. \ref{fig:3} (right). The $FUV-r$ C09 models are not plotted - difference in metallicity has little effect on this colour for C09 at $z<1$  and these colours track those of the low $Y$ YEPS models with an offset due to the different values of the RGB mass loss parameter ($\eta$) used in each case. The photometry for the $z=0.55$ and $z=0.68$ clusters (observed through the F275W filter) needed to be transformed to rest-frame  $1550$\AA. This transformation was minimal for the SDSSJ1004+4112 photometry (from 1650 to 1550\AA) and  larger for the two lower redshift clusters (from 1750 to 1550\AA). In order to carry out this transformation, we linearly interpolated the range in $1550-1650$ and $1550-1750$ from the range of $FUV-NUV$ colours displayed by the YEPS models, as well as upturn galaxies in low redshift clusters (typically $1<FUV-NUV<2$). We chose a correction appropriate to $FUV-NUV=1.5$ and added error bars reflecting the full range and therefore maximum uncertainty in potential colours, combined with the measured photometric uncertainties.

Fig. \ref{fig:yeps_fuv} (right) shows the same photometric results as Fig. \ref{fig:yeps_fuv} (left) but now plotted against the evolution of the $FUV-r$ colours as predicted by the binary model of \cite{han2007} and the metal-rich and metal-poor models from \cite{yi1999}. The lack of evolution to $z=0.55$ is clearly inconsistent with the predictions of the low metallicity model, as the upturn only develops about 2 Gyrs ago in this case. Similarly, the detection of a transition in the presence of upturns at $z=0.7$ falsifies the binary evolution model of \cite{han2007}, as such a transition is not expected to be observed, except at very high redshift. Finally, increasing mass loss as a function of metallicity is also less appealing as an explanation, as there is no reason for the rise of the upturn to take place at any particular redshift, and there is also evidence that mass loss on the RGB is not dependent on metal abundance \citep{miglio2012,salaris2016}, while the upturn is observed to correlate with metallicity in early-type galaxies (\citealt{burstein1988}). Of the commonly discussed models that seek to explain the upturn and specifically those discussed in this and our previous papers, our observations therefore  appear to support only the hypothesis that the upturn derives from He-rich HB stars, where the transition in upturn strength is a primary prediction of that model.

It is also very clear from this plot that even the two detected galaxies in SDSSJ1004+4112 are redder in $FUV-r$ than the extremes of the range seen in all lower redshift clusters, and the photometry of the non (or barely) detected stack, which contains the combined light from  the majority of the red sequence galaxies around $L^*$, is consistent with little to no upturn. This is completely inconsistent with the results from the lower redshift clusters - the 5$\sigma$ limit from the SDSSJ1004+4112 stack is as red or redder than the reddest galaxies in all the other clusters. The only way that this cluster shows similar behaviour to those at lower redshift is that its BCG has a similar strength upturn to the other BCGs, perhaps indicating the different fundamental properties of BCGs relative to the rest of the cluster early-type population.

As noted in our previous work, these results also have implications for the minimum stellar mass of the galaxies at the earliest stages in their evolution. The results presented here and interpreted in terms of a He-enhanced population origin for the upturn imply that the upturn population is formed by $z=4$ at the latest {\it in situ} (see also \citealt{goudfrooij2018}). Our work in paper I implies that this population accounts for several percent (and potentially up to $\sim 20$ per cent in the most extreme objects) of the total stellar mass of each galaxy. For the most massive of the cluster red sequence galaxies today this implies  minimum stellar masses at $z=4$ of order $10^{10}$ M$_{\odot}$.  This value is almost certainly  even larger if, as seems plausible,  any He-rich subpopulation is produced from a previous stellar generation (with yields as in globular clusters, the initial generation of stars providing the enrichment must be about 20 times more massive than the second generation -- e.g. \citealt{dantona2016}). Such massive objects are difficult to produce within CDM simulations.

All of the above discussion assumes a single value for metallicity and formation redshift for the YEPS models, but in Fig. \ref{fig:8} we show the effect of different metallicities and formation redshifts on the YEPS model with $Y=0.42$ (from Fig. \ref{fig:yeps_fuv}) as an example. As can be seen from the plots, while the onset of the upturn does depend on these parameters, it is clear that for the models to fit the observed behaviour of the upturn, early formation redshifts ($z_{f}\geqslant4$) and comparatively high metallicities (i.e. solar or above) are generally required, supporting our choice of $z_{f}$ and metallicity in the above discussion.

\subsection{Alternative explanations and caveats}
The hot HB populations in second parameter globular clusters are generally believed to be He-rich, from both direct and indirect evidence. Our observations (in this and previous papers) imply that such hot HB stars are also present in some numbers in early-type galaxies. It is reasonable to assume that these are again  He-rich, as this is consistent with the evidence we have presented and with observations of comparatively metal-rich bulge globular clusters in our Galaxy (e.g. \citealt{piotto2005,piotto2007}). \cite{goudfrooij2018} suggests that these He-rich stars were formed within now dissolved metal-rich globular clusters, whose FUV-bright counterparts are observed around M87 and other galaxies (\citealt{peacock2017}). In order to achieve the observed $FUV-V$ colours we require high He abundances, especially given the high metallicities of early-type galaxies; these are comparable to the more extreme stars in Omega Centauri, NGC2808, NGC6388 and NGC6441. The origin of the extra Helium is still in dispute for the Milky Way Globular cluster systems, and even more so for galaxies (\citealt{karakas2006}; \citealt{maeder2006}; \citealt{cassisi2009}; \citealt{moehler2011}; \citealt{ventura2013}; \citealt{chantereau2016}). Massive (3-5M$_{\odot}$) AGB stars (\citealt{denissenkov1997}) or Fast Rotating Massive Stars (\citealt{denissenkov2011,denissenkov2013}) have been proposed as candidate polluters, although neither explains the detailed abundance patterns and both require a somewhat contrived star formation history (see review by \citealt{bastian2018}). However, the balance of observational evidence favours a He-rich subpopulation as the most likely explanation (see papers I and II), even if the origin of the Helium pollution is uncertain.

There are several other parameters that can potentially affect the $NUV-r$ colour of a galaxy besides the presence of an upturn stellar population, particularly the metallicity and age of the main sequence population, which could in theory account for the observed spread in the $NUV-r$ colour. In paper II we explained in detail why these parameters cannot be the strongest driving factors behind the scatter in $NUV-r$, and that it is the upturn that dominates this colour in red sequence galaxies. Here we summarise the key points in our arguments (see paper II for further details). By comparing our results with C09 models of solar, sub-solar and super-solar metallicities as seen in Fig. \ref{fig:yeps}, we found that even a comparatively low metallicity model (0.56\(Z_\odot\)), despite being unrealistic for most of our giant elliptical/S0 population around the $L^*$ point (which are likely to have Z=1-2\(Z_\odot\) as shown for Coma by \citealt{price2011}), still cannot account for the full range in $NUV-r$ colours shown by our galaxies. Furthermore, as can be seen from any of the C09 models in Fig. \ref{fig:yeps}, a change in age of the SSP by 1 Gyr only brings about a change of 0.1 mags in $NUV-r$. Given that most cluster red sequences seem to be established by $z\sim2$ (\citealt{newman2014}; \citealt{glazebrook2017}), and most star-formation in these galaxies seems to have stopped before $z=2$ (\citealt{kodama1998}; \citealt{jorgensen2017}), our reasonable estimate of $z_f$=4 can only shift by approximately 1 Gyr, clearly insufficient to account for the observed spread in $NUV-r$. As such, an extra component of upturn is required to explain the full spread in $NUV-r$ seen in all of our clusters.

One final parameter that can affect our estimate of the age and $Y$ of the stellar populations from the YEPS models is the Reimers' mass loss parameter - $\eta$ (\citealt{reimers1975,reimers1977}). Once again the effect of this parameter is discussed in greater detail in paper II, but in summary, the choice of the mass loss parameter can directly affect the strength and onset of the upturn (since a higher mass loss on the RGB leads to a higher surface temperature of the star in the HB). The $\eta$ parameter primarily affects the $FUV-NUV$ colour, giving rise to the offset between the $FUV-r$ colours in the C09 and YEPS $Y=0.27$ (non-upturn) models as noted earlier. The YEPS models plotted in Fig. \ref{fig:yeps} assume $\eta=0.63$ as calibrated using Milky Way globular clusters, which is also reasonable for the majority of our galaxies. But changing the value of this parameter could likely bring about an uncertainty of $\sim1$ Gyr in the age of our models.

Finally, we note here the main caveats in our results. SDSSJ1004+4112 is only one cluster for which the decline in the upturn has been observed in the general red sequence population; suitable data does not yet exist for other clusters at the same or higher redshift. Although SDSSJ1004+4112 is an optically poorer system (in terms of the number of red sequence galaxies selected by us) than many of the other clusters studied in this series of papers, it is still a relatively massive cluster ($\sim 3 \times 10^{14} M_{\odot}$ -- \citealt{oguri2012} ) showing significant strong and weak gravitational lensing. Our comparison of $FUV-V$ colours for galaxies in Coma, Perseus and Fornax - which themselves span a decade in mass -  shows that this colour does not depend on environment. We have also used 24 clusters at $z < 0.12$ from 2dFGRS where we have measured $FUV$ and $NUV$ colours for a complete sample of spectroscopically identified members down to $M_K=-21$, with SDSS and PanStarrs1 optical colours. These are discussed in detail in a forthcoming paper. We find that the $FUV-r$ colour of their red sequence galaxies does not depend on cluster velocity dispersion or X-ray luminosity, and therefore our results in SDSSJ1004+4112 are likely to apply to the general population of galaxies at its redshift. In other words, the upturn phenomenon is internal to galaxies, and therefore unrelated to star formation history as affected by the cluster environment. A similar result was obtained by \cite{boissier2018}, \cite{yi2011} and \cite{loubser2011}, who also suggested that the upturn is intrinsic to galaxies and not related to their environments. 

While the fading in the upturn is therefore very likely to take place at around $z=0.7$ given the limited time required to evolve a He-enhanced hot HB prior to this, it is possible that SDSSJ1004+4112 may be an outlier given that it is not as rich in high mass red sequence galaxies as the two $z=0.55$ systems we compare it to. To fully confirm our results, it would be best to observe several more clusters in the UV at $z=0.7-1$. This small redshift window is particularly important as it is during this time that the upturn is likely to be developing while  the observed epoch is late enough that the bulk of the population is passively evolving (and therefore comparatively red). Regardless of the observed fading in the upturn at $z=0.7$, which places an upper limit on the amount of He-enhancement required to produce the observed upturn, a $Y$ of 0.45 (or higher for earlier formation redshifts than $z_f$=4) is necessary to account for the upturn seen in MACSJ1149+2223 and MACSJ0717+3745 at $z=0.55$, along with all other clusters at lower redshifts.

\section{Conclusions}

We have measured the evolution of the UV upturn to $z=0.7$ from archival UV images of four clusters. We detect no evolution in the strength and range of upturn exhibited by cluster red sequence galaxies out to $z=0.55$ but then observe a  strong decline to $z=0.7$. This behaviour rules out most of the theoretical models for the origin of the upturn, but is predicted by those  where the increased UV emission arises from a population of He-enhanced stars which have evolved onto the (hot) horizontal branch. This implies that a fraction of the stellar population  in such galaxies (perhaps up to $\sim 10-20\%$ for the those with the strongest upturns)  has high ($Y > 0.45$) Helium abundance and large formation ages ($z_f > 4$), and that objects at these redshifts had {\it in situ} stellar masses of the order of $10^{10}\ M_{\odot}$ at these early epochs.

\section*{Acknowledgements}

SSA is funded by an STFC PhD studentship and thanks the University of Turku and FINCA for their hospitality and local funding during the visits where part of this work was carried out.  This work was based on observations made with the NASA/ESA Hubble Space Telescope, obtained from the data archive at the Space Telescope Science Institute. STScI is operated by the Association of Universities for Research in Astronomy, Inc. under NASA contract NAS 5-26555.
We acknowledge discussions with Alessandro Bressan on this topic.


\bibliographystyle{mnras}
\bibliography{references}


\label{lastpage}
\end{document}